\documentclass{llncs}
\usepackage{algorithmic}
\usepackage{graphicx,amssymb,amstext,amsmath}
\usepackage{subfigure}
\usepackage{epstopdf}
\usepackage{float}
\usepackage{url}
\usepackage{verbatim}
\usepackage{xspace}
\begin{document}
\title{Model Checking Tap Withdrawal in C. Elegans}
\author{Md. Ariful Islam\inst{1}
\and Richard DeFrancisco\inst{1}
\and Chuchu Fan\inst{3}
\and Radu Grosu\inst{1,2}
\and Sayan Mitra\inst{3}
\and Scott A. Smolka\inst{1} 
}
\institute{Stony Brook University
\and Vienna University of Technology
\and University of Illinois Urbana Champaign
}
\maketitle

We present what we believe to be the first formal verification of
a biologically realistic (nonlinear ODE) model of a neural circuit
in a multicellular organism: Tap Withdrawal (TW) in \emph{C.~Elegans},
the common roundworm.  TW is a reflexive behavior exhibited by
\emph{C.~Elegans} in response to vibrating the surface on which it
is moving; the neural circuit underlying this response is the
subject of this investigation.  Specifically, we perform reachability
analysis on the TW circuit model of Wicks et al.\ (1996), which 
enables us to estimate key circuit parameters.  Underlying our
approach is the use of Fan and Mitra's recently developed technique
for automatically computing local discrepancy (convergence and
divergence rates) of general nonlinear systems.  We show that the
results we obtain are in agreement with the experimental results of
Wicks et al.\ (1995).  As opposed to the fixed parameters found in
most biological models, which can only produce the predominant behavior,
our techniques characterize ranges of parameters that produce (and do
not produce) all three observed behaviors: reversal of movement,
acceleration, and lack of response.

\section{Introduction}
\label{sec:introduction}

Although neurology and brain modeling/simulation is a popular
field of biological study, formal verification has yet to take
root.  There has been cursory study into neurological model
checking (see Section~\ref{sec:related_work}), but not with the
nonlinear ODE models used by biologists.  We believe that the
insights gained through formal verification and analysis can
transform the field, as has been the case in the
\emph{Electronic Design Automation} (EDA) industry, which is
now valued at over \$4 billion annually.  As EDA has allowed for
increased complexity for a smaller time investment in hardware
circuits, we believe that the same kind of benefits can be
realized for neural circuits.

For our initial neurological study, we have selected the round worm,
\emph{Caenorhabditis Elegans}, due to the simplicity of its nervous
system (302 neurons, $\sim$5,000 synapses) and the breadth of research
on the animal.  The complete connectome of the worm is documented, and
there have been a number of interesting experiments on its response to
stimuli.

For model-checking purposes, we were particularly interested in the
\emph{tap withdrawal} (TW) neural circuit.  The TW circuit governs the
reactionary motion of the animal when the petri dish in which it swims
is perturbed.  (A related circuit, \emph{touch sensitivity}, controls
the reaction of the worm when a stimulus is applied to a single point
on the body.)  Studies of the TW circuit have traditionally involved
using lasers to ablate the different neurons in the circuit of multiple
animals and measuring the results when stimuli are applied.
%% Many of these studies use slightly different circuits or derive
%% slightly different mathematical models.

A model of the TW circuit was presented by Wicks, Roehrig, and Rankin
in~\cite{Wicks96}.  This model is in the form of a system of nonlinear
ODEs, as well as mapped polarities of the various neurons involved
in the TW circuit.  Additionally, Wicks and Rankin had a previous
paper in which they measure the three possible reactions of the
animals to TW with various neurons ablated~\cite{Wicks95}; see
also Fig.~\ref{fig:bar_graph}.  The three behaviors---acceleration,
reversal of movement, and no response---are logged with the
percentage of the experimental population to display that behavior.

The~\cite{Wicks96} model has a number of circuit parameters, such as
gap-junction conductance, capacitance, and leakage current, that
crucially affect the behavior of the organism.  Values for these
parameters based on estimates, rules of thumb and measurements are
given in~\cite{Wicks96}, but not the parameter ranges.  A quick
analysis of the circuit shows that variations in these parameter
values may give rise to changes in the behavior of the model from
acceleration or reversal to no-response.  As all biological parameters
vary  across populations, time, and environments, identifying parameter
ranges corresponding to behaviors is a fundamentally important problem. 
%% (Here we may want to cite Donze's work on parameter synthesis,
%% see below)
For a complete characterization of the TW circuit, it is therefore
critical to identify the range of parameter values that
give rise to these different types of behavior.

Using automatically generated local
\emph{discrepancy functions}~\cite{FanM:2015,DMVemsoft2013}, we are
able to perform reachability analysis on the~\cite{Wicks96} model.
This approach combines static analysis with numerical simulations to 
allow us to iteratively compute more precise over-approximations of
the reachable states of the system with respect to a continuous range
of parameter values.  We used this approach, which we refer to as
$\delta$-\emph{refinement}, to determine parameter ranges that produce
all three behaviors for the control group (no ablation) and four
ablation experiments.  This is a significant expansion of the biological
results, where only static parameters are obtained, and only for one
behavior per experimental group.  The specific parameters of interest are
the gap-junction conductances for three sensory neurons in the TW circuit,
as the gap junctions formed by these neurons are known to be the most
important functional connections to the TW process.

Our results are further organized into how many of these conductances we
considered simultaneously, a categorization we refer to as 1-D, 2-D, and
3-D.  For the 1-D and 2-D cases, we were able to determine parameter
ranges for which the three TW responses are guaranteed to hold.  The
3-D case is only applicable to the control group; here, again, we were
able to produce the same guarantees.  Moreover, with a single exception
(which Wicks himself has experienced), our results match the trends
(in terms of relative percentages) of the earlier biological
experiments (see Fig.~\ref{fig:bar_graph}).

The rest of the paper develops along the following lines.
Section~\ref{sec:background} provides requisite background material
on the TW neural circuit, its reactionary behavior, and the ODE model
of~\cite{Wicks96}.  Section~\ref{sec:approach} describes our 
reach-tube reachability analysis and associated property checking.
Section~\ref{sec:results} presents our extensive collection of
model-checking/parameter-estimation results.
Section~\ref{sec:related_work} reviews related work.
Section~\ref{sec:conclusion} offers our concluding remarks and
directions for future work.

\section{Related Work}
\label{sec:related_work}

Iyengar et al.~\cite{2007Nettab} present a Pathway Logic (PL) model of 
neural circuits in the marine mollusk \emph{Aplysia}.  Specifically,
the circuits they focus on are those involved in neural plasticity and
memory formation.  PL systems do not use differential equations,
favoring qualitative symbolic models.  They do not argue that they can
replace traditional ODE systems, but rather that their qualitative
insights can support the quantitative analysis of such systems.  
Neurons are expressed in terms of rewrite rules and data types.  Using 
the PL formalism, they are able to simulate neural circuits and perform 
qualitative \emph{in silico} experiments, such as simulating knock-out
of an individual components or other changes to the network.
Their simulations, unlike our reachability analysis, do not provide 
exhaustive exploration of the state space.  Additionally, PL models
are abstractions usually made in collaboration between computer 
scientists and biologists.  Our work meets the biologists on their own 
terms, using the pre-existing ODE systems developed from physiological 
experiments.

Tiwari and Talcott~\cite{2008Tiwari} build a discrete symbolic model of 
the neural circuit Central Pattern Generator (CPG) in \emph{Aplysia}.
The CPG governs rhythmic foregut motion as the mollusk feeds.  Working
from a physiological (non-linear ODE) model, they abstract to a discrete
system and use the Symbolic Analysis Laboratory (SAL) model checker to
verify various properties of this system.  They cite the complexity of
the original model and the difficulty of parameter estimation as
motivation for their abstraction.  Their discrete model synchronously
composes 10 input/output automaton (neurons), connects them with 3 types
of links (excitatory synapse, inhibitory synapse, gap-junction), and
includes an observer component.  The input of each neuron can be positive, 
negative, or zero and the output is a boolean: a pulse is generated or
not.  Our  approach uses the original biological model of the TW circuit
of \emph{C. Elegans}~\cite{Wicks96}, and through reachability analysis,
we obtain the parameter ranges of interest.

We have extensive experience with model checking and reachability
analysis in the cardiac domain, e.g.~\cite{cav11,HuangFMMK14,hscc14,hscc15}.
In fact, much of our previous work has focused on the cardiac myocyte, a 
computationally similar cell to the neuron.  This is not surprising
as both belong to the class of \emph{excitable cells}.  The similarities
are so numerous that we have used a variation of the Hodgkin-Huxley
model of the squid giant axon~\cite{hh} to model ion channel flow in
cardiac tissue.

\section{Background}
\label{sec:background}

%In this section, we describe the Tap Withdrawal (TW) neural circuit
%of \textit{C. Elegans}.  We then discuss a mathematical model of
%the circuit.

% Biological Background
%\subsection{The Tap Withdrawal Circuit}

In \textit{C. Elegans}, there are three classes of neurons:
\emph{sensory}, \emph{inter}, and \emph{motor}.  For the TW circuit,
the sensory neurons are \emph{PLM}, \emph{PVD}, \emph{ALM,} and
\emph{AVM}, and the inter-neurons are \emph{AVD}, \emph{DVA},
\emph{PVC}, \emph{AVA}, and \emph{AVB}.  The model we are using
abstracts away the motor neurons as simply forward and reverse
movement.

\begin{figure}[h!]
    \centering
    \includegraphics[scale=0.45]{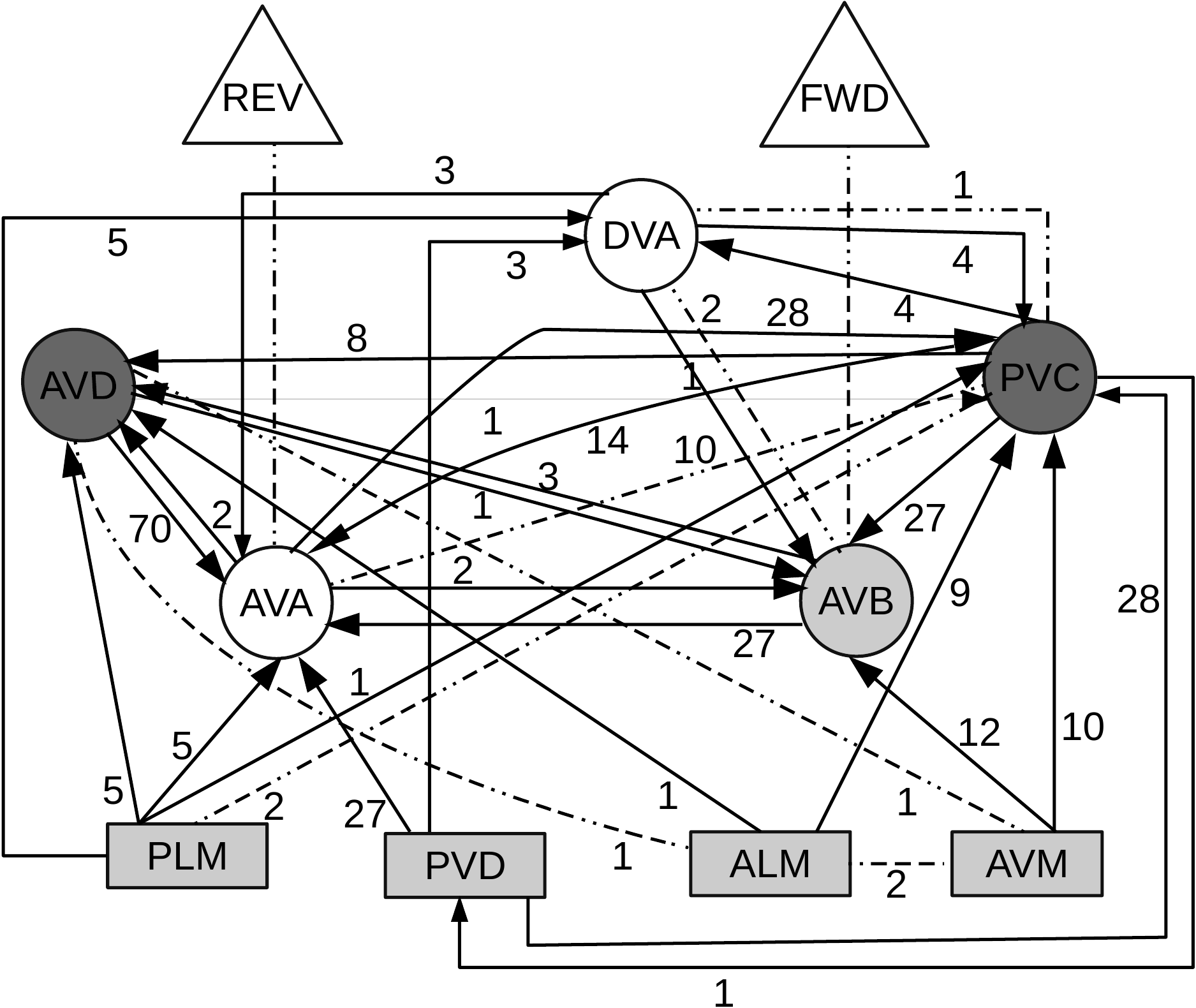}
\caption{Tap Withdrawal Circuit of C. Elegans. Rectangle: Sensory
Neurons; Circle: Inter-neurons; Dashed Undirected Edge: Gap Junction;
Solid Directed Edge: Chemical Synapse; Edge Label: Number of
Connections; Dark Gray: Excitatory Neuron; Light Gray: Inhibitory
Neuron; White: Unknown Polarity.  FWD: Forward Motor system; REV:
Reverse Motor System.}
%% the motor systems are not part of the tap withdrawal circuit.
\label{fig:TWCircuit}
\end{figure}

Neurons are connected in two ways: electrically via bi-directional
\emph{gap junctions}, and chemically via uni-directional chemical
\emph{synapses}.  Each connection has varying degrees of throughput,
and each neuron can be \emph{excitatory} or \emph{inhibitory},
governing the polarity of transmitted signals.  These polarities
were experimentally determined in~\cite{Wicks96}, and used to
produce the circuit shown in Fig.~\ref{fig:TWCircuit}.  

\begin{figure}[h!]
    \centering
    \includegraphics[scale=0.5]{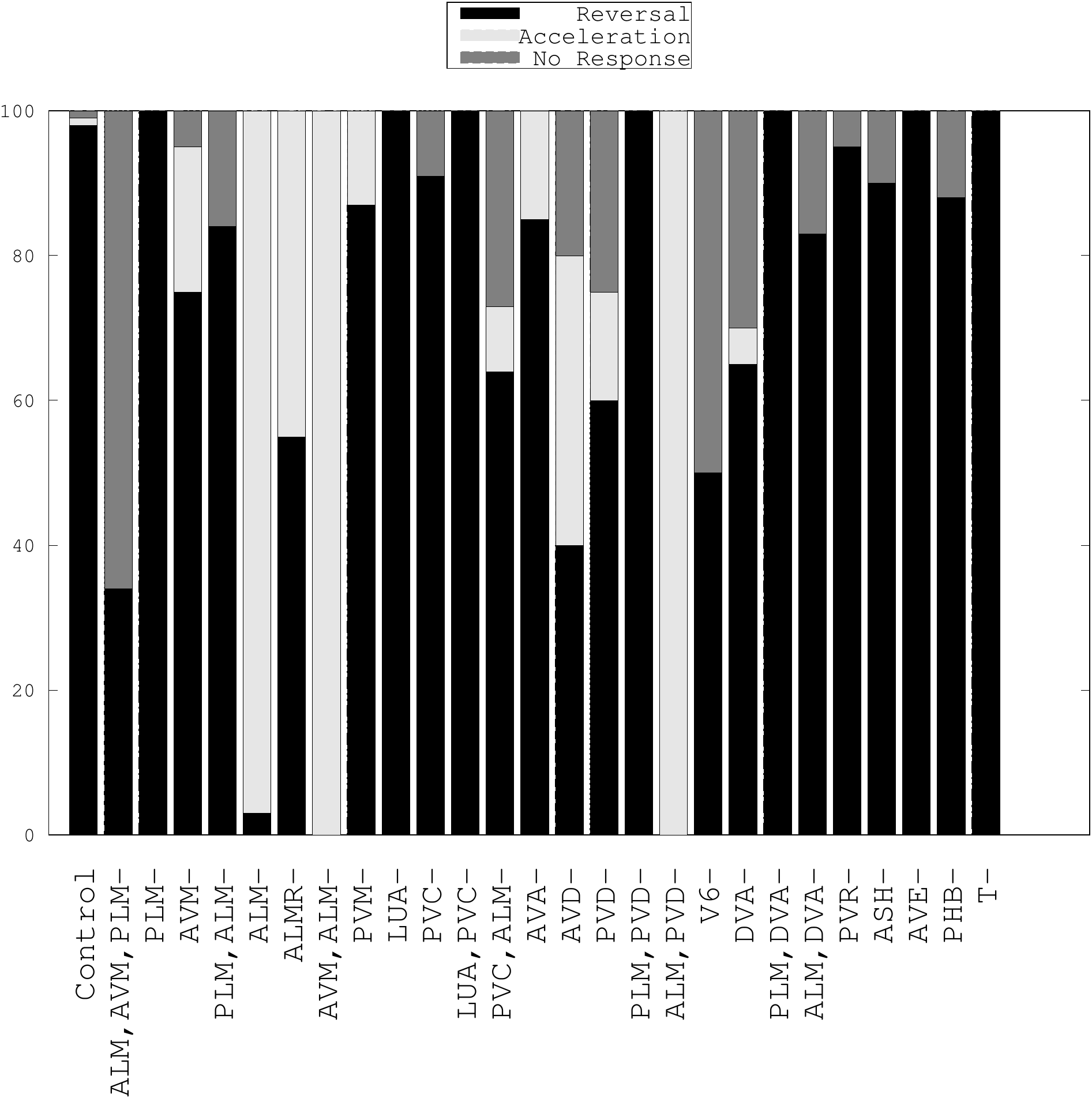}
\caption{Effect of ablation on Tap Withdrawal reflex.  The length 
of the bars indicate the fraction of the population demonstrating the 
particular behavior.}
\label{fig:bar_graph}
\end{figure}

The TW circuit produces three distinct locomotive behaviors:
\emph{acceleration}, \emph{reversal} of movement, and a
\emph{lack of response}.  In~\cite{Wicks95}, Wicks et al. performed
a series of laser ablation experiments in which they knocked out a
neuron in a group of animals (worms), subjected them to a tapped
surface, and recorded the magnitude and direction of the resulting
behavior.  In the control group with no neurons knocked out, 98\% of
subjects reacted to a tap with a reversal of locomotion, but there
were still measured cases of acceleration and ``no response'' behavior.
Fig.~\ref{fig:bar_graph} shows the response types for each of their
experiments.

%

% Mathematical Background
%\subsection{Mathematical Model for the Tap Withdrawal Circuit}

%In~\cite{Wicks96}, Wicks et al.\ presented a mathematical model
%for the TW circuit.  This subsection provides an overview of this
%model.

%\subsection*{Modeling the Neuron}
The dynamics of a neuron's membrane potential, \emph{V}, is
determined by the sum of all input currents, written as:
\begin{equation*}
C_m\dot{V}=\frac{1}{R_m}(V_l - V)+\sum{I^{gap}}+\sum{I^{syn}}+I^{stim}
\end{equation*}
where $C_m$ is the \emph{membrane capacitance}, $R_m$ is the
\emph{membrane resistance}, $V_l$ is the \emph{leakage potential},
$I^{gap}$ and $I^{syn}$ are gap-junction and the chemical 
synapse currents, respectively, and $I^{stim}$ is the applied
external \emph{stimulus} current.  The summations are over all 
neurons with which this neuron has a (gap-junction or synaptic) connection.
%

%
%No Space for this figure right now
%
%\begin{figure}[h!]
%\label{fig:neuron_dynamics}
%    \centering
%    \includegraphics[scale=0.3]{figures/neuron_dynamics.pdf}
%\caption{Schematic diagram of neuron dynamics of Neuron $i$.}
%\end{figure}
%

%\subsection*{Modeling Gap-Junction Currents}
The current flow between neuron $i$ and $j$ via a gap-junction
is given by:
\begin{equation*}
I_{ij}^{gap} = n^{gap}_{ij}g^{gap}_m(V_j-V_i)
\end{equation*}
where the constant $g^{gap}_m$ is the \emph{maximum conductance}
of the gap junction, and $n^{gap}_{ij}$ is the number of
gap-junction connections between neurons $i$ and $j$.  The conductance 
$g^{gap}_m$ is one of the key circuit parameters of this model that dramatically 
affects the behavior of the animal.
%

%\subsection*{Modeling Chemical Synapse Currents}
The synaptic current flowing from pre-synaptic neuron $j$ to
post-synaptic neuron $i$ is described as follows:
\begin{equation*}
I_{ij}^{syn} = n^{syn}_{ij}g^{syn}_{ij}(t)(E_j - V_i)
\end{equation*} 
where $g^{syn}_{ij}(t)$ is the time-varying synaptic conductance of
neuron $i$, $n^{syn}_{ij}$ is the number of synaptic connections
from neuron $j$ to neuron $i$,  and $E_j$ is the
\emph{reversal potential} of neuron $j$ for the synaptic conductance. 

The chemical synapse is characterized by a synaptic sign, or polarity,
specifying if said synapse is excitatory or inhibitory.  The value of
$E_j$ is assumed to be constant for the same synaptic sign; its value
is higher if the synapse is excitatory rather than inhibitory. 

Synaptic conductance is dependent only upon the membrane potential of
presynaptic neuron $V_j$, given by:
\begin{equation*}
g^{syn}_{ij}(t) = g^{syn}_{\infty}(V_j)
\end{equation*}
where $g^{syn}_{\infty}$ is the steady-state
\emph{post-synaptic conductance} in response to a pre-synaptic
membrane potential.

The steady-state post-synaptic membrane conductance is 
modeled as:
\begin{equation*}
g^{syn}_\infty(V_j) =
\frac{g^{syn}_m}{1+\exp{(-4.3944\frac{V_j-V_{EQ_j}}{V_{Range}})}}
\end{equation*}
where $g^{syn}_m$ is the
\emph{maximum post-synaptic membrane conductance} for the synapse,
$V_{EQ_j}$ is the \emph{pre-synaptic equilibrium potential}, and
$V_{Range}$ is the \emph{pre-synaptic voltage range} over which the
synapse is activated.

%
%\subsection*{Modeling the Tap Withdrawal Circuit}
Combining all of the above pieces, the mathematical model of the 
TW circuit is a system of nonlinear ODEs, with each state variable 
defined as the membrane potential ofa neuron in the circuit.  Consider a 
circuit with $N$ neurons.  The dynamics of the $i^{th}$ neuron of the circuit
is given by:
\begin{equation}
\label{eq:circuit_dyn_1}
C_{m_i}\dot{V_i}=\frac{V_{l_i} -
  V_i}{R_{m_i}}+\sum_{j=1}^N{I_{ij}^{gap}}+\sum_{j=1}^N{I_{ij}^{syn}}
+I^{stim}_i
\end{equation}
\begin{equation}
\label{eq:circuit_dyn_2}
I^{gap}_{ij}=n^{gap}_{ij}g^{gap}_m(V_j-V_i)
\end{equation}
\begin{equation}
\label{eq:circuit_dyn_3}
I^{syn}_{ij}=n^{syn}_{ij}g^{syn}_{ij}(E_{j}-V_i)
\end{equation}
\begin{equation}
\label{eq:circuit_dyn_4}
g^{syn}_{ij} =
\frac{g^{syn}_m}{1+\exp{(-4.3944\frac{V_j-V_{EQ_j}}{V_{Range}})}}.
\end{equation}

The equilibrium potentials ($V_{EQ}$) of the neurons are computed by setting
the left-hand side of Eq.~(\ref{eq:circuit_dyn_1}) to zero.  This
leads to a system of linear equations, that can be solved as follows:
\begin{equation}
V_{EQ}=A^{-1}b
 \label{eq:Linear}
\end{equation}
where matrix $A$ is given by:
\begin{equation*}
A_{ij}=
\begin{cases}
-R_{m_i}n^{gap}_{ij}g^{gap}_m & \text{if } i \neq j \\
1+R_{m_i}\sum_{j=1}^N{n^{gap}_{ij}g^{gap}_{ij}g^{syn}_m/2} & \text{if } i=j
\end{cases}
\end{equation*}
and vector $b$ is written as:
\begin{equation*}
b_i = V_{l_i}+R_{m_i}\sum_{j=1}^N{E_jn^{syn}_{ij}g^{syn}_m/2}.
\end{equation*}
%

%\subsection*{Output of the Tap Withdrawal Circuit} 
The potential of the motor neurons \emph{AVB} and \emph{AVA}
determine the observable behavior of the animal.  If the integral of 
the difference between $V_{\mathit{AVA}}$ - $V_{\mathit{AVB}}$ is large, the animal will 
reverse movement.  By extension, if the difference is a large negative 
value, the animal will accelerate, and if the difference is close to zero 
there will be no response.  The equation that converts the membrane 
potential of \emph{AVB} and \emph{AVA} to a behavioral property, 
(e.g. \emph{reversal}), is given by:
\begin{equation}
\text{Propensity to Reverse} \propto \int (V_{\mathit{AVA}} - V_{\mathit{AVB}}) dt
 \label{eq:Reversal}
\end{equation}  
where the integration is computed from the beginning of tap
stimulation until either the simulation ends or the integrand
changes sign.  To allow initial transients after the tap,
the test for a change of integrand sign occurs only after a grace
period of 100~ms.

%\subsection{Normalization of Tap Withdrawal Circuit Model}

For the purpose of reachability analysis
(Section~\ref{sec:approach}), we normalize the system of equations with
respect to the capacitance.  Combining Eqs.(~\ref{eq:circuit_dyn_1})
and (~\ref{eq:circuit_dyn_4}) and taking $C_{m_i}$ to the right-hand
side, we have:

\scriptsize
\begin{multline*}
%\label{eq:circuit_dyn_norm}
\dot{V_i}=\frac{V_{l_i} - V_i}{R_{m_i} C_{m_i}}+
\frac{g^{gap}_m}{C_{m_i}}\sum_{j=1}^N{n^{gap}_{ij}(V_j-V_i)}
+\frac{g^{syn}_m}{C_{m_i}}\sum_{j=1}^N{\frac{n^{syn}_{ij}(E_{j}-V_i)}
{1+\exp{(-4.3944\frac{V_j-V_{EQ_j}}{V_{Range}})}}} 
+\frac{1}{C_{m_i}}I^{stim}_i
\end{multline*}
\normalsize

Now letting $g^{leak}_i = \frac{1}{R_{m_i} C_{m_i}}$,
$g^{gap}_i = \frac{g^{gap}_m}{C_{m_i}}$,
$g^{syn}_i = \frac{g^{syn}_m}{C_{m_i}}$ and $I^{ext}_i = \frac{1}{C_{m_i}}$
the system dynamics can be written as:
\scriptsize
\begin{multline}
\label{eq:circuit_dyn_norm}
\dot{V_i}=g^{leak}_i (V_{l_i} - V_i)+
g^{gap}_i \sum_{j=1}^N{n^{gap}_{ij}(V_j-V_i)}
+g^{syn}_i \sum_{j=1}^N{\frac{n^{syn}_{ij}(E_{j}-V_i)}
{1+\exp{(-4.3944\frac{V_j-V_{EQ_j}}{V_{Range}})}}} 
+I^{ext}_i
\end{multline}
\normalsize
This is the 9 dimensional ODE model of the TW circuit.  The key 
circuit parameters are the gap conductances, $g_i^{gap}$, and we 
aim to characterize the ranges of these conductances that 
produce acceleration, reversal, and no response.

\newcommand{\num}[1]{\relax\ifmmode \mathbb #1\else $\mathbb #1$\fi}
\newcommand{\nnnum}[1]{\relax\ifmmode
  {\mathbb #1}_{\geq 0} \else ${\mathbb #1}_{\geq 0}$
  \fi}
\newcommand{\npnum}[1]{\relax\ifmmode
  {\mathbb #1}_{\leq 0} \else ${\mathbb #1}_{\leq 0}$
  \fi}
\newcommand{\pnum}[1]{\relax\ifmmode
  {\mathbb #1}_{> 0} \else ${\mathbb #1}_{> 0}$
  \fi}
\newcommand{\nnum}[1]{\relax\ifmmode
  {\mathbb #1}_{< 0} \else ${\mathbb #1}_{< 0}$
  \fi}
\newcommand{\plnum}[1]{\relax\ifmmode
  {\mathbb #1}_{+} \else ${\mathbb #1}_{+}$
  \fi}
\newcommand{\nenum}[1]{\relax\ifmmode
  {\mathbb #1}_{-} \else ${\mathbb #1}_{-}$
  \fi}
\newcommand{\unsafe}{{\sf U}}                    %unsafe
\newcommand{\reals}{{\num R}}                    %reals
\newcommand{\booleans}{{\num B}}                    %reals
\newcommand{\nnreals}{{\nnnum R}}                    %nonnegative reals
\newcommand{\realsinfty}{{\num R} \cup \{\infty, -\infty\}}
%nonnegative reals
\newcommand{\plreals}{{\plnum R}}                    %positive reals
\newcommand{\naturals}{{\num N}}                      %natural numbers
\newcommand{\integers}{{\num Z}}                      %integers
\newcommand{\rationals}{{\num Q}}                      %rationals
\newcommand{\nnrationals}{{\nnnum Q}}                   % nonnegative rationals
\newcommand{\Time}{{\num T}}
\newcommand{\reach}[1]{\relax\ifmmode {\sf Reach}_{#1} \else ${\sf
    Reach}_{#1}$\fi}
\def\C{{\cal C}} % HA
%%%%%%%%%%%
\section{Reachability Analysis of Nonlinear TW Circuit}
\label{sec:approach}
% Sayan: I assume the connection between reachability and parameter range
% estimation 
% has been introduced at this point.
Reachability analysis for verifying properties for general nonlinear
dynamical systems is a well-known hard problem.  Our approach relies
on a recent line of investigation that combines static analysis of the
model with validated numerical
simulations~\cite{DMVemsoft2013,HuangFMMK14,C2E2:TACAS2015}.

\subsection{Background on Reachability using Discrepancy}
Consider an $n$-dimensional {\em autonomous dynamical system}:
\begin{eqnarray} \label{eq:dynamic system}
\dot{x} = f(x),
\label{eq:system}
\end{eqnarray}
where $f:\reals^n \rightarrow \reals^n$ is a Lipschitz continuous
function. A \emph{solution} or a \emph{trajectory} of the system is
a function $\xi:\reals^n \times \nnreals \rightarrow \reals^n$ such
that for any initial point $x_0 \in \reals^n$ and at any time $t >0$,
$\xi(x_0, t)$ satisfies the differential equation~(\ref{eq:system}).
A state $x$ in $\reals^n$ is {\em reachable from the initial set $\Theta
  \subseteq \reals^n$ within a time interval $[t_1, t_2]$\/} if there exists an
  initial state $x_0 \in \Theta$ and a time $t \in [t_1,t_2]$ such that $x
  =\xi(x_0,t)$. The set of all reachable states in the interval $[t_1,t_2]$ is
  denoted by $\reach{}(\Theta,[t_1,t_2])$. If $t_1 =0$, we write
  $\reach{}(t_2)$ when set $\Theta$ is clear from the context.
If we can compute or approximate the reach set of such a model, then
we can check for invariant or temporal properties of the model.
Specifically, \emph{C. Elegans} TW properties such as accelerated forward
movement or reversal of movement fall into these categories.
Our core reachability algorithm~\cite{DMVemsoft2013,HuangFMMK14,C2E2:TACAS2015}
uses a simulation engine that gives sampled numerical simulations
of~(\ref{eq:system}).
%
% define simulation
\begin{definition} \label{def:simulation}
A {\em $(x_0, \tau, \epsilon,T)$-simulation\/} of~(\ref{eq:system}) is a
sequence of time-stamped sets  $(R_0, t_0)$, $(R_1,t_1) \ldots, (R_n,t_n)$
satisfying:
\begin{enumerate}
\item Each $R_i$ is a compact set in $\reals^n$ with $\mathit{dia}(R_i) \leq
  \epsilon$.
\item The last time $t_n = T$ and for each $i$, $0 < t_i - t_{i-1} \leq \tau$,
  where the parameter $\tau$ is  called the {\em sampling period\/}.
\item For each $t_i$, the trajectory from $x_0$ at $t_i$ is in $R_i$, i.e.,
  $\xi(x_0,t_i) \in R_i$, and 
for any $t \in [t_{i-1}, t_i]$, the solution $\xi(x_0,t) \in
  hull(R_{i-1},R_i)$.
\end{enumerate}
\end{definition}
%Simulation engines generate a sequence of states and error bounds using
%  numerical integration.
%Libraries like CAPD~\cite{capd} and VNODE-LP~\cite{vnode2006} compute such
%simulations
%for a wide range of nonlinear dynamical system models and 
%the $R_i$'s are represented by some data structure like hyperrectangles.
% Sayan: Maybe drop this def.
%For a set of states $D\subseteq \reals^n$, a {\em $(D, \tau, T)$-reachtube\/}
%of~(\ref{eq:system}) is a sequence of time-stamped sets $(R_0, 0), (R_1,t_1)
%\ldots, %(R_n,t_n)$ satisfying:
%\begin{enumerate}
%\item Each $R_i \subseteq \reals^n$ is a compact set of states.
%\item The last time $t_n = T$ and for each $i$, $0 \leq t_i - t_{i-1} \leq
%\tau$.
%\item For any $x_0 \in D$, and any time $t \in [t_{i-1}, t_i]$, the solution
%$\xi(x_0,t) \in R_i$.
%\end{enumerate}
The algorithm for reachability analysis uses a key property of the model
called a {\em discrepancy function\/}.
\begin{definition}
\label{def:disc}
A uniformly continuous function $\beta:\reals^n \times \reals^n \times \nnreals
\rightarrow \nnreals$ is a {\em discrepancy function\/} of~(\ref{eq:system}) if
\begin{enumerate}
\item for any pair of states $x, x' \in \reals^n$, and any time $t >0$,
\begin{eqnarray}
\|\xi(x,t) - \xi(x',t)\| \leq \beta(x,x',t), \mbox{and}
\label{eq:df1}
\end{eqnarray}
\item for any $t$, as $x \rightarrow x'$, $\beta(.,.,t) \rightarrow 0$.
\end{enumerate}
\end{definition}
If a function $\beta$ meets the two conditions for any pair of states $x,x'$ in
a compact set $K$ then it is called a {\em $K$-local discrepancy function.}
Uniform continuity means that $\forall \epsilon>0,  \forall x,x' \in K, \exists
\delta$ such that for any time $t, \|x-x'\|<\delta \Rightarrow
\beta(x,x',t)<\epsilon.$
The verification results
in~\cite{DMVemsoft2013,HuangFMMK14,DuggiralaWMVM14,C2E2:TACAS2015} required
the user to provide the discrepancy function $\beta$ as an additional
input for the model.
A Lipschitz constant of the dynamic function $f$ gives an exponentially growing
$\beta$, contraction metrics~\cite{ContractionNonlinear} can give tighter bounds for
incrementally stable models, and sensitivity analysis gives tight bounds for
linear systems~\cite{donze2007systematic}, but none of these give an
algorithm for computing $\beta$ for general nonlinear models. Therefore,
finding the discrepancy can be a barrier in the verification of large models
like the TW circuit.

Here, we use Fan and Mitra's recently developed
approach that automatically computes local discrepancy along individual
trajectories~\cite{FanM:2015}. 
Using the simulations and discrepancy, the reachability algorithm for checking
properties proceeds as follows:
Let the $\unsafe$ be the set of states that violate the invariant in question.
% Here is the source
First, a $\delta$-cover $\C$ of the initial set $\Theta$ is computed; that is,
the union of all the $\delta$-balls around the points in $\C$ contain $\Theta$.
This $\delta$ is chosen to be large enough so that the cardinality of $\C$
is small.  Then the algorithm iteratively and selectively refines
$\C$ and computes more and more precise 
over-approximations of $\reach{}(\Theta,T)$ as a union $\cup_{x_0 \in \C}
\reach{}(B_\delta(x_0),T)$.
Here, $\reach{}(B_\delta(x_0),T)$ is computed by first generating a 
$(x_0, \tau, \epsilon,T)$-simulation and then bloating it by a factor
that maximizes $\beta(x,x',t)$ over $x,x' \in B_\delta(s_0)$ and $t \in
[t_{i-1},t_i]$.
If $\reach{}(B_\delta(x_0),T)$ is disjoint from $\unsafe$ or is (partly)
contained in $\unsafe$, 
then the algorithm decides that $B_\delta(x_0)$ satisfies and violates
$\unsafe$, respectively. 
Otherwise, a  finer cover of $B_\delta(x_0)$ is added to $\C$ and the iterative
selective refinement continues.  We refer to this in this paper as $\delta$-refinement.
In~\cite{DMVemsoft2013}, it is shown that this algorithm is sound and
relatively complete for proving bounded time invariants. 
% that is, if the system is robustly safe, the algorithm will terminate and
% return ``SAFE''. If any executions from $\Theta$ is unsafe, it will terminate
% and return % ``UNSAFE''.
%
%The procedure first generates a set of numerical approximations of the
%behaviors from a finite sampling of the initial states. 
%% TAP specific
%% SM: I am using bf temporarily to highlight statements that are TAP specific.
%
%Bloating each of these simulations by an appropriately large factor we compute
%an over-approximation of the reach set. 
%
%For soundness, the bloating factor should be chosen to be large enough to make
%each bloated simulation an over approximation of the reachable states of the
%system from a neighborhood of the initial state. 
%
%On the other hand, for relative completeness (modulo the precision of the
%machine), it should be possible to make the error due to bloating arbitrarily
%small for any point in time.
%
%These two opposing requirements are captured in the definition of a {\em
%discrepancy function\/} of~\cite{DMVemsoft2013}: For an $n$-dimensional
%dynamical system, it is any function $\beta:\reals^{2n} \times \nnreals
%\rightarrow \nnreals$, such that (a) it gives an upper-bound on the distance
%between any two trajectories $\xi(x,t)$  and $\xi'(x,t)$ of the system
%$|\xi(x,t) - \xi(x',t)| \leq \beta(x,x',t)$, and (b) it vanishes as $x$
%approaches $x'$.  
%
\subsection{Applying Local Discrepancy to TW Circuit}
Fan and Mitra's algorithm (see details in~\cite{FanM:2015}) for automatically
computing local discrepancy relies on the Lipschitz constant and the
Jacobian of
the dynamic function, along with simulations. The Lipschitz constant is used to
construct a coarse, one-step over-approximation $S$ of the reach set of the
system along a simulation. Then the algorithm computes an upper bound on the
maximum eigenvalue of the symmetric part of the Jacobian over $S$, using a
theorem from matrix perturbation theory. This gives a piecewise exponential
$\beta$, but the exponents are tight as they are obtained from the maximum
eigenvalue of the linear approximation of the system in $S$. This means that
for models with convergent trajectories, the exponent of $\beta$ over $S$ will
be negative, and the $\reach{}(T)$ approximation will quickly become very
accurate.
In the rest of this section, we describe key steps involved in making this
approach work with the TW circuit. 
% computing L Lipchitz constant 
% computing Jacobian

The model of the TW circuit from Section~\ref{sec:background} can be
written as $\dot V =
f(V)$, where $V \in \reals^9$.
The Jacobian of the system is the matrix of partial derivatives
with the $ij^{th}$ term given by:

\scriptsize

\begin{eqnarray}
\frac{\partial f_i}{\partial V_i} &=& -g_i^{leak}-g_i^{gap}\sum_{j=1,j\neq
  i}^{N}n_{ij}^{gap}-g_{i}^{syn}\sum_{j=1,j\neq
  i}^{N}\frac{n_{ij}^{syn}}{1+\exp(-4.3944\frac{V_j-V_{EQ_{j}}}{V_{Range}})}
  \nonumber \\
 &=&
  g_i^{gap}n_{ij}^{gap}-g_{i}^{syn}n_{ij}^{syn}\frac{\frac{-4.3944}{V_{Range}}\exp(-4.3944\frac{V_j-V_{EQ_{j}}}{V_{Range}})(E_j-V_i)}{(1+\exp(-4.3944\frac{V_j-V_{EQ_{j}}}{V_{Range}}))^2}
  \label{eq:TWjacobian}
\end{eqnarray}

\normalsize

For parameter-range estimation of the TW circuit, each parameter $p$ of
interest is added as a new variable with constant dynamics ($\dot{p} =
0$). Computing the reach-set from initial values of $p$ is then used
to verify or falsify invariant properties for a continuous range of parameter
values, and therefore a whole family of models, instead of analyzing just a
single member of that family. 
Here the parameters of interest are the quantities
$1/g_i^{leak},10/g_i^{gap},1/g_i^{syn}$. 
%{\bf SM2AI: requires some bio-inspired justification.}
%
%The total number is 0~27 depends on the parameter we want to check the range. 
Consider, for example, $1/g_i^{leak}$ as a
parameter: 
\scriptsize
%\[
%\dot{\left[
%    \begin{array}{c}
%       V \\ 
%        1/g_i^{leak}
%	\end{array}
%ne
% 0 & 0 & 0 & 0
% \end{array}
%\right].
%\] 
\[
\dot{\left[
	\begin{array}{c}
		 V \\ 
		 1/g_i^{leak}
		\end{array}
		\right]}
		= 
		\left[
			\begin{array}{c}
				f(V) \\
				0
			\end{array}
			\right].
\] 
\normalsize
%{\bf SM2CF: So, what what should we say now that we have shown that this can
%be computed? interesting observations can we make about this? Did we have to
%do any further engineering to make this work? How did you choose the time
%horizon for coordinate transformation?\/}  
In this case the Jacobian matrices for the system with
parameters will be singular because of the all-zero rows that come from the
parameter dynamics. The zero eigenvalues of these singular matrices are taken
into account automatically by the algorithm for computing local discrepancy.  In this 
paper we focus on $10/g_i^{gap}$, leaving the others for future work.
%
%{\bf SM2SS: There is one more engineering detail that we can talk about
%  here. It is related to coordinate transformation that the algorithm performs
%  internally. I doubt if it sharpens the overall narrative though.\/}
\subsection{Checking Properties}
\label{sec:props}
Once the reach sets are computed, checking the \textit{acceleration},
\textit{reversal}, and \textit{no-response} properties are conceptually
straightforward.
%{\bf SM2AI: lets standardize the property names across the paper.}
For instance, Equation~(\ref{eq:Reversal}) gives a method to check reversal movement. 
Instead of computing the integral of $({V_{\mathit{AVA}} - V_{\mathit{AVB}}})$, we use the
  following sufficient condition to check it:
\[
\forall \ t \in T_{\mathit{int}}, \forall \ x \in \reach{}(\Theta, [t,t]),
V_{\mathit{AVA}}(x) > V_{\mathit{AVB}(x)}.
\]
Here, $T_{int}$ is a specific time interval after the stimulation time, 
$\Theta$ is the initial set with parameter ranges, 
and recall that $\reach{}(\Theta, [t,t])$ is the set of states reached
at time $t$ from $\Theta$.
We implement this check by scanning the entire reachtube and checking that its
projection on $V_{\mathit{AVB}}(x)$
is above that of $V_{\mathit{AVA}}(x)$ over all intervals. If this check
succeeds (as in Figure~\ref{fig:propCheck}(a)),
we conclude that the range of parameter values produce the reversal movement. 
If the check fails, then the reversal
movement is not provably satisfied (Figure~\ref{fig:propCheck}(b))
and in that case we $\delta$-refine the initial partition. 
%
%% commented out by Scott:  In our implementation, we only perform a few
%% refinements.
%{\bf SM2AI: modify last part of this section to lead to experiments. }
\begin{figure}[h!]
\centering
\subfigure[\scriptsize Rev.\ property satisfied with $g^{gap}_{AVM}=1000$.]
{\includegraphics[height=45mm,width=58mm]
{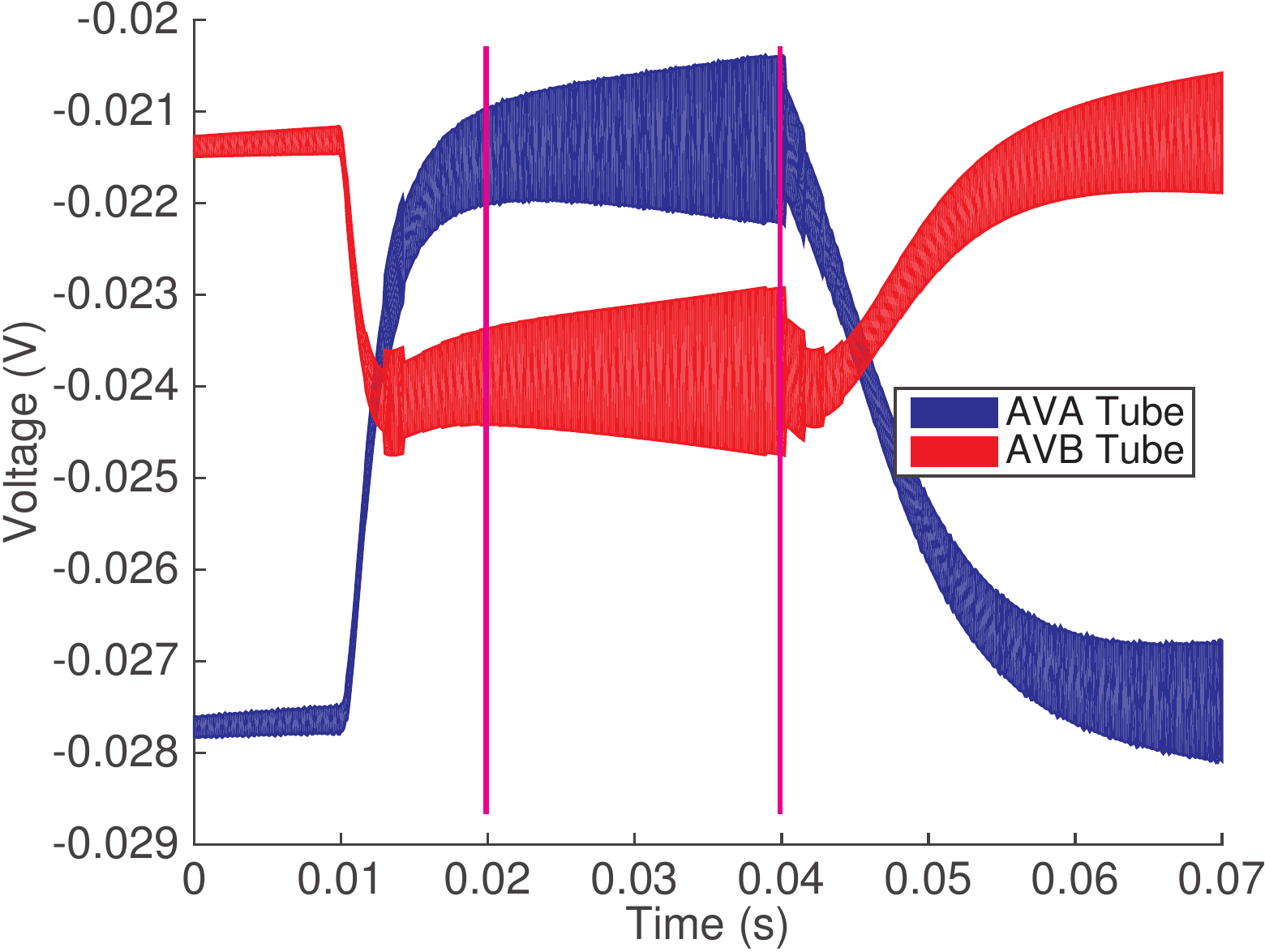}}
\quad
\subfigure[\scriptsize Rev.\ unknown with $g^{gap}_{AVM}=33.33$.]
{\includegraphics[height=45mm,width=58mm]
{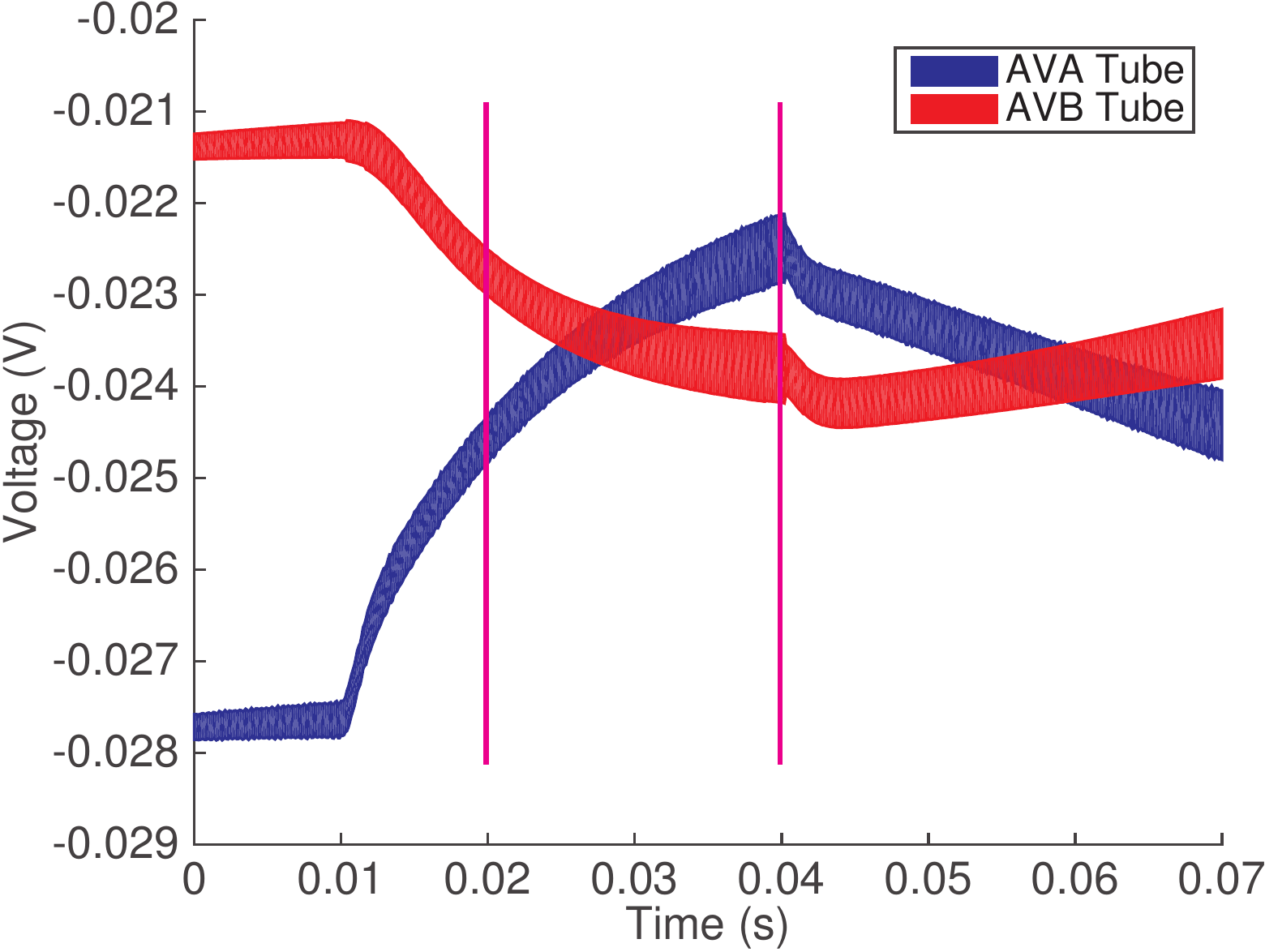}}
\caption{\scriptsize Model Checking Reversal Property of Control Group, with
  $\delta=5e-5$, varying  
 $g^{gap}_{\mathit{AVM}}$.}
\label{fig:propCheck}
\end{figure}
\begin{figure}[h!]
\centering
\subfigure[\scriptsize Rev.\ property unknown with $\delta=1e$-$4$.]
{\includegraphics[height=45mm,width=58mm]
{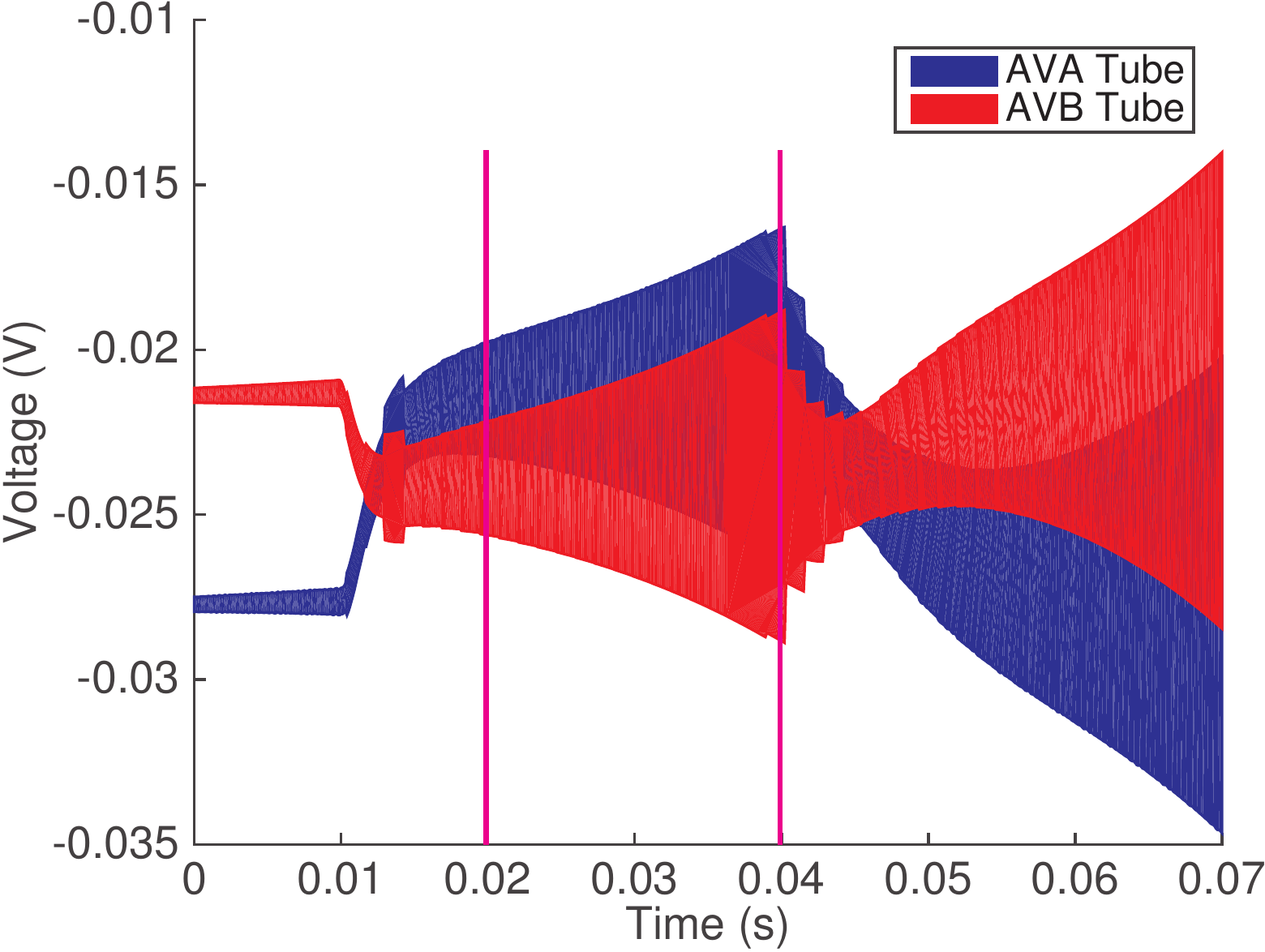}}
\quad
\subfigure[\scriptsize Rev.\ property satisfied with $\delta=5e$-$5$.]
{\includegraphics[height=45mm,width=58mm]
{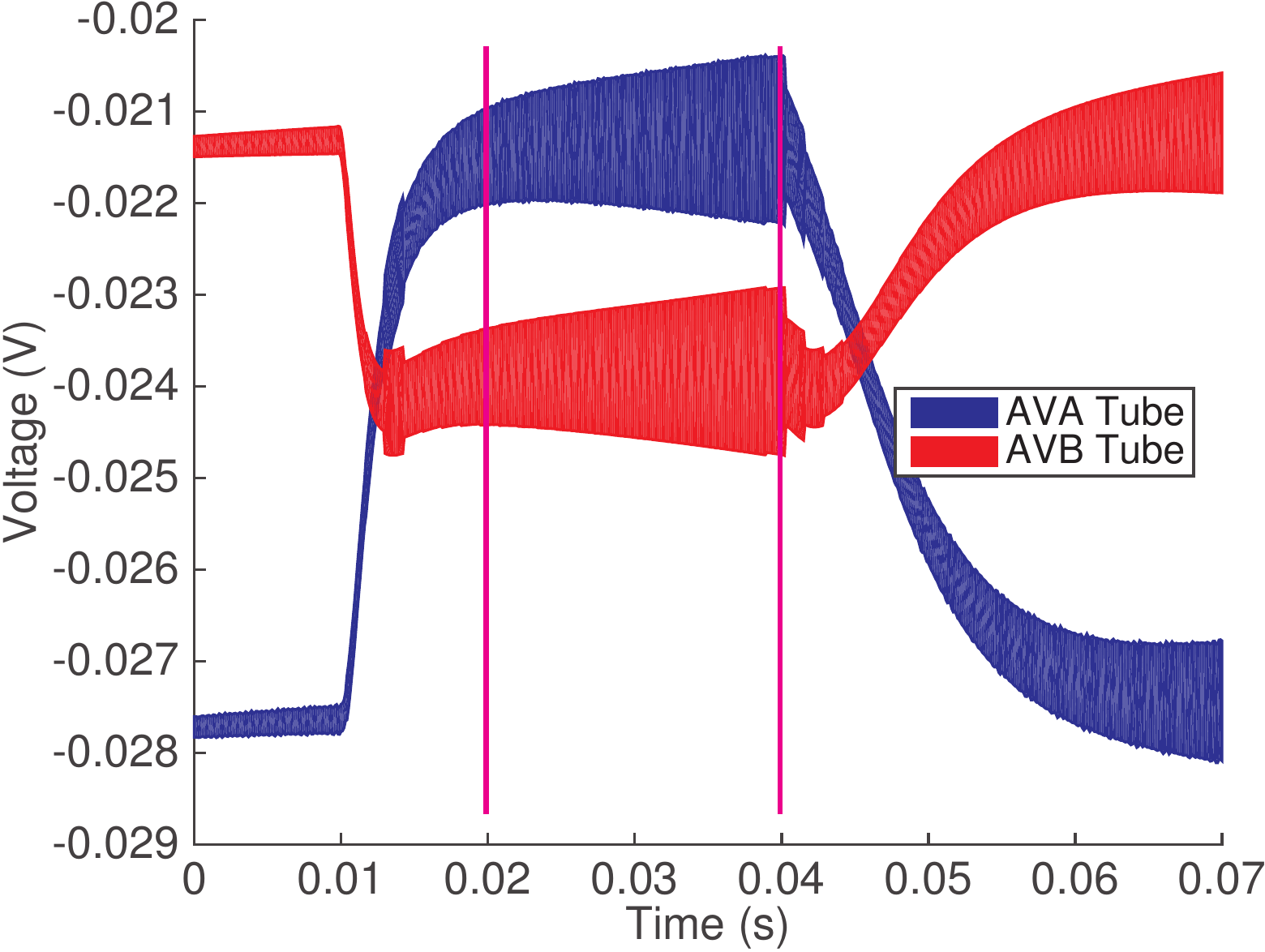}}
\caption{\scriptsize Model Checking Reversal Property of Control Group by
  refining $\delta$. }
\label{fig:propCheck}
\end{figure}
%Chuchu: As for linear transformations, can we point to the section of our
% paper? Basically we just do some coordinate transformation to make the
% over-approximation less conservative.
%For example,
%Consider a simple linear system:  
%\begin{equation}\label{linear transformation example}
%\dot x = \left[ \begin{array}{cc}
  %0 & 3 \\
  %-1& 0
%\end{array} 
%\right]x,
%\end{equation}
%
%the Jacobian matrix is constant, and the discrepancy function without
% coordinate transformation is:
%\[
%\|\xi (x_{1},t)-\xi (x_{2},t)\| \leq \|x_{1}-x_{2}\|e^{t-t_1}.
%\]
%If we use $P = \left[ \begin{array}{cc}
%1 & 3 \\
%-\sqrt{3} & \sqrt{3} \\
%\end{array} \right]$ as the coordinate transformation matrix, $\tilde J =
%PJP^{-1} = 
%\left[ \begin{array}{cc}
%0 & \sqrt{3} \\
%-\sqrt{3} & 0 \\
%\end{array} \right],$
%and the discrepancy function with coordinate transformation is 
%\[
%\|\xi (x_{1},t)-\xi (x_{2},t)\| \leq \sqrt{3}\|x_{1}-x_{2}\|.
%\]

\section{Experimental Results}
\label{sec:results}
In this section, we apply reachability analysis to parameter rangers that produce 
three different behaviors (\textit{reversal}, \textit{acceleration}, \textit{no response}) in the control 
and four ablation groups.  When a tap stimulus is applied, the sensory neurons (\textit{ALM}, \textit{AVM} 
and \textit{PLM})  propagate that signal to the motor neurons via interneurons.  The  
gap-junctions formed by the sensory neurons are the most important functional connections to this 
process~\cite{Chalfie85}.  Therefore we vary only the  gap-junction conductance, $g^{gap}_i$, 
of the sensory neurons and keep all other parameters constant, as per~\cite{Wicks96}. 
Our experiments characterize parameter ranges for \textit{reversal}, \textit{acceleration} and \textit{no response}
behaviors in all groups, except the ALM,AVM- group where \textit{reversal} behavior is not observed.  
%We notice that if we set $g^{gap}_i$  for other neurons to the same value used in~\cite{Wicks96}, 
%the TW circuit will always produce a response.  So to produce \textit{no response} behavior 
%we additionally need to reduce the $g^{gap}_i$ of other neurons to a small value ($1000$ times smaller 
%than that of the one in~\cite{Wicks96}).  

In section~\ref{sec:approach}, 
we explain that we use $10/g^{gap}_i$ as our parameter in the state vector instead of $g^{gap}_i$.  
Assume $p^{gap}_{i}=10/g^{gap}_i$, $i\in\{AVM, ALM, PLM\}$.  The corresponding range for $p^{gap}_i$ is
$[0.01,1]$.  From the reachability analysis, we estimate ranges for $p^{gap}_i$ that can be
converted back to $g^{gap}_i$.

In the following subsections, we will present our results for parameter range estimation for all three  behaviors of 
the control and ablation groups.  This process requires three experiments per group.

\subsection{1-D Parameter Space}
Here we vary one conductance at a time for two groups: the control and the \textit{ALM}, \textit{AVM} ablation
groups.

\subsubsection{Control:}
For the control case, we varied $p^{gap}_{\mathit{AVM}}$.  We found that the \textit{reversal} 
property is satisfied in sub-range $[0.01,0.214]$ with $\delta=1e-6$ and the \textit{acceleration} 
property in sub-range $[0.63,1]$ with $\delta=1e-5$.  Recall, $\delta$ is the size of the finest cover 
used by the verification algorithm.  We could not verify any property for the 
sample points in sub-range $(0.214,0.63)$.  As shown in Table~\ref{table:Conductance},  the 
parameter range producing \textit{reversal}, as identified by our procedure, dominates the 
parameter range for \textit{acceleration}.  Our procedure also shows that no value of $p^{gap}_{AVM}$ 
produces the no-response behavior for the control group.

The time required for our procedure is dependent upon the property, the interval for each dimension, and the 
size of $\delta$.  For example, the time necessary to complete the procedure for the \textit{reversal} property 
is approximately one hour.

\subsubsection{ALM, AVM Ablation Group:}
In this group two sensory neurons, \textit{ALM}, and \textit{AVM}, are ablated.
As such, we vary only $p^{gap}_{\mathit{PLM}}$.  
\textit{Acceleration} is satisfied over the interval $[0.01,0.3]$ with $\delta=5e-5$ 
and \textit{no response} behavior is satisfied over $[0.75,1]$ with the same $\delta$.  
Despite using a very
small $\delta$ for refinement, we did not observe any \textit{reversal} behavior in this entire range.
Examining Table~\ref{table:Conductance} we see that \textit{acceleration} is the dominant behavior for this group.

\subsection{2-D Parameter Space}
We lead with results for the control group, then examine various ablation groups.

\subsubsection{Parameter Refinement in 2-D:}
\begin{figure}[h!]
\centering
{\includegraphics[width=100mm]
{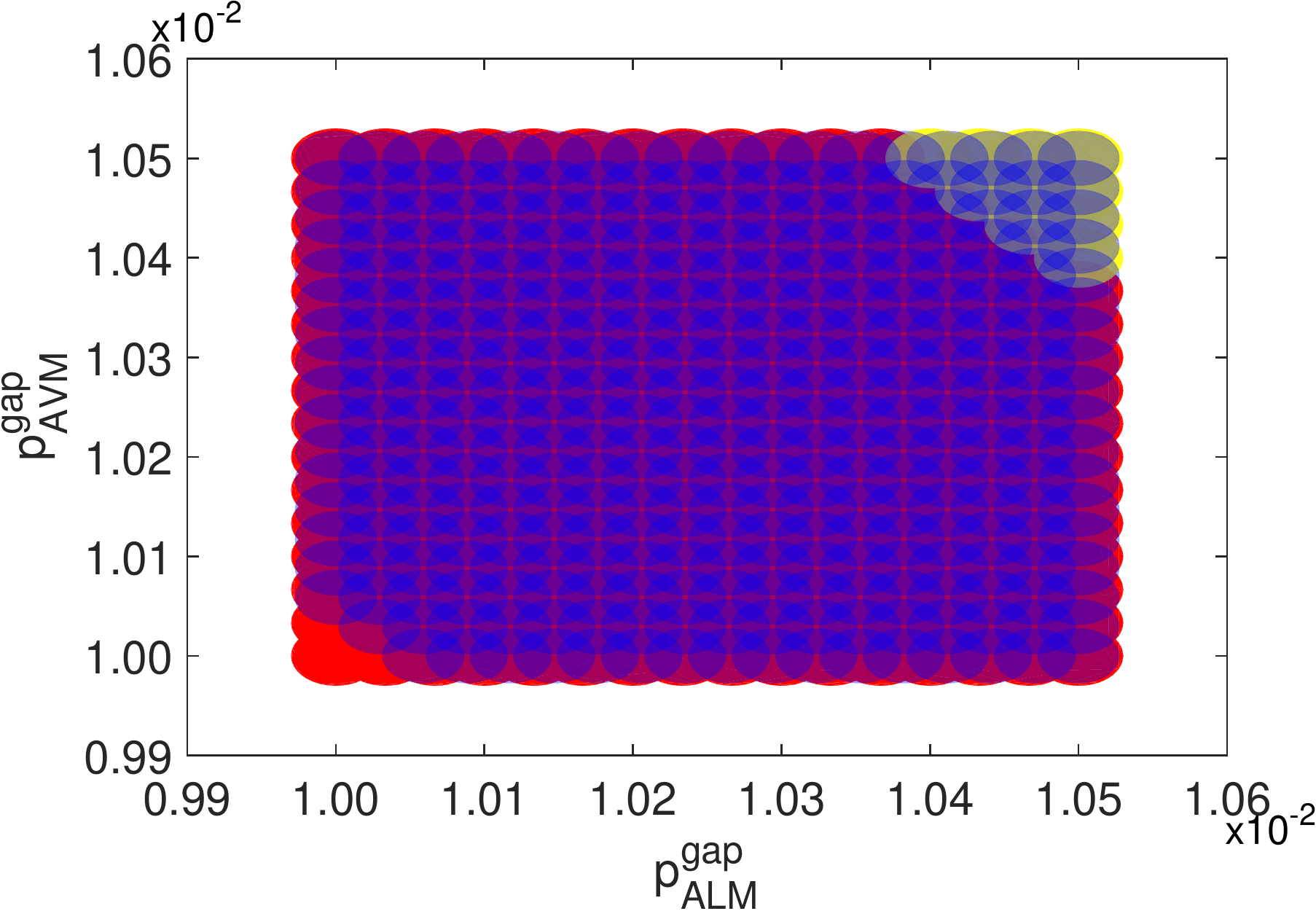}}
\caption{Example of 2-D Parameter Refinement.  Red Regions are Unknown for both 
$\delta=6e-5$ and $\delta=5.5e-5$, Red/Blue Regions are Unknown for $\delta=6e-5$, but Satisfied for 
$\delta=5.5e-5$, and Yellow/Blue Regions are Satisfied for both.}
\label{fig:refinement}
\end{figure}

Fig.~\ref{fig:refinement} helps paint a picture of how the $\delta$-refinement process discussed in 
Section~\ref{sec:approach} works in 2-D.  We consider 4 refinement steps for the control group: 
$\delta=7e-5$, $\delta=6e-5$, $\delta=5.5e-5$, and $\delta=5e-5$.  For $\delta=7e-5$, the property of interest is unknown at 
all points.  With $\delta=6e-5$ the property is considered unknown for all red areas in the figure, 
including red and blue areas.  Blue areas show where $\delta=5.5e-5$ are satisfied, and in the 
blue and yellow area both $\delta=6e-5$ and $\delta=5.5e-5$ have a satisfied property.  The property 
is satisfied for the entire range of the graph when $\delta=5e-5$.  Thus, the refinement process stops at 
$\delta=5e-5$, and the entire range of the 2-D parameter space is characterized.

\subsubsection{Control}
Here we consider the $p^{gap}_{\mathit{AVM}}$ and $p^{gap}_{\mathit{ALM}}$ conductances simultaneously.  
For this group, \textit{reversal} is satisfied over the range $[0.01,0.0105]$ with $\delta=2e-5$ 
and \textit{acceleration} is satisfied over $[0.63,0.6305]$ with the same $\delta$.  Table~\ref{table:Conductance} 
shows that \textit{reversal}, like in the 1-D case, dominates and \textit{no response} is not generated.

%7B paragraph  
%For the \textit{reversal} property, we sampled 10 points uniformly from the range [0.01, 0.214] for 
%both $p^{gap}_{AVM}$ and $p^{gap}_{ALM}$.  Then we checked the property on the computed 
%reach-tubes for 100 (10x10) sample points in the two dimensional parameter space.  Fig.~\ref{fig:control_rev}(a) 
%shows the bigger, but sampled parameter space in 2-D case, where the reversal property is 
%satisfied. Even though this figure does not cover the space completely, the clear demarcation 
%line suggests that, taking large enough samples, we might be able to compute a much bigger 
%space than what we computed. We notice the similar phenomena for other properties in all groups.

% 
%\begin{figure}[h!]
%\centering
%
%7B
%\subfigure[\scriptsize 2-D Case, with $\delta=5e-6$.]
%{\includegraphics[width=58mm]
%{figures/control_rev_2d_5e-6.pdf}}
%9B
%\subfigure[\scriptsize 3-D Case, with $\delta=1e-6$.]
%{\includegraphics[width=58mm]
%{figures/control_rev_3d_1e-6.pdf}}
%\caption{ Model Checking Reversal 
%Property of Control Group in
%sampled space.  Blue and Red Denote property Satisfied
%and Unknown, respectively.}
%\label{fig:control_rev}
%\end{figure}
%

% 
\begin{table}[h!]
\centering
{\includegraphics[width=120mm]
{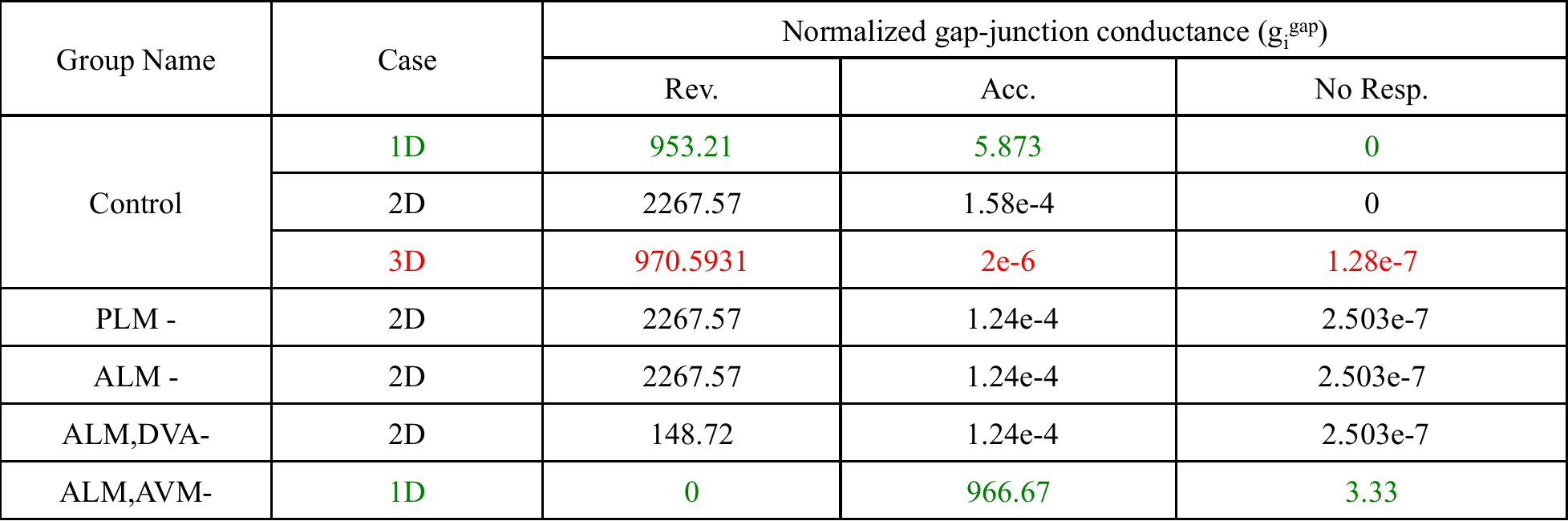}}
\caption{Regions in the parameter space in which 
the properties are proven satisfied.  A 1-D case 
shows interval size, a 2-D case shows area, and a 3-D case shows 
volume.}
\label{table:Conductance}
\end{table}

\subsubsection{PLM Ablation Group:}
As the \textit{PLM} neuron is ablated in this group, varying only $p^{gap}_{\mathit{AVM}}$ and $p^{gap}_{\mathit{ALM}}$
is sufficient to produce all three behaviors.  Here we find \textit{reversal} satisfied over 
$[0.01,0.0105]$ with $\delta=2e-5$, \textit{acceleration} over $[0.67,0.6705]$ with $\delta=5e-5$, and 
\textit{no response} over $[0.9995,1]$ with $\delta=5e-5$.  Table~\ref{table:Conductance} shows that 
\textit{reversal} dominates the other two behaviors, but all three are produced.

\subsubsection{ALM Ablation Group:}
To produce all three behaviors of this group we vary only $p^{gap}_{\mathit{AVM}}$ and $p^{gap}_{\mathit{PLM}}$.  
\textit{Reversal} is satisfied over the interval $[0.01,0.0105]$ with $\delta=5e-5$, \textit{acceleration}
 over $[0.67,0.6705]$ with $\delta=2e-5$, and \textit{no response} over $[0.9995,1]$ with $\delta=5e-5$.
We can see in Table~\ref{table:Conductance} that this ablation group has a propensity to \textit{reverse}.  
The astute reader would notice that this trend does not seem to match Fig.~\ref{fig:bar_graph}, unlike 
the rest of our results.  We have run simulations with the equations from ~\cite{Wicks96}, and the simulations 
also produce \textit{reversal}, not \textit{acceleration}.  The results of the simulation and model checking 
consistently disagree with the behavior denoted in Fig.~\ref{fig:bar_graph} for this \textit{ALM} group.  
We are currently investigating why this is the case.

\subsubsection{ALM, DVA Ablation Group:}
All three behaviors of this group are produced by varying only $p^{gap}_{\mathit{AVM}}$ and $p^{gap}_{\mathit{PLM}}$.  
Here \textit{reversal} is satisfied over $[0.02,0.0205]$ with $\delta=2e-5$, \textit{acceleration} over 
$[0.67,0.6705]$ with $\delta=5e-5$, and \textit{no response} over $[0.9995,1]$ with $\delta=5e-5$.
Repeating our experiments for this group, Table~\ref{table:Conductance} shows the dominant \textit{reversal} 
behavior.

\subsection{3-D Parameter Space}
Since the ablation groups we have used in this paper all feature at least one of the primary sensory 
neurons (\textit{ALM}, \textit{AVM}, and \textit{PLM}) ablated, we can only show the 3-D case for 
the original animal.

For the 3-D case, in addition to $p^{gap}_{\mathit{AVM}}$ and $p^{gap}_{\mathit{ALM}}$, we have 
the $p^{gap}_{\mathit{PLM}}$ conductance.   Finally, we get a non-zero 
value for \textit{no response} in the control, but Table~\ref{table:Conductance} shows that this 
value is an order of magnitude smaller than \textit{acceleration} and several orders 
smaller than \textit{reversal}.  \textit{Reversal} is satisfied over 
$[0.01,0.0101]$ with $\delta=2e-5$, \textit{acceleration} over $[0.631,0.6305]$ with $\delta=5e-5$, and 
\textit{no response} over $[0.63,0.63005]$ with $\delta=5e-5$.

%Below this is cut

\section{Conclusions}
\label{sec:conclusion}

In this paper, we performed reachability analysis with discrepancy to
automatically determine parameter ranges for three fundamental
reactions by \emph{C. Elegans} to tap-withdrawal stimulation:
reversal of movement, acceleration, and no response.  We followed the
lead of the \emph{in vivo} experimental results of~\cite{Wicks95} to
obtain parameter-estimation results for gap-junction conductances for
a number of neural-ablation groups.
%% including the control group for which none of the animal's neurons
%% are ablated.
To the best of our knowledge, these results represent the first formal
verification of a biologically realistic (nonlinear ODE) model of a
neural circuit in a multicellular organism.

Our results are further organized into how many of these three
conductances we considered simultaneously.   For each of these cases,
we were able to determine parameter ranges for which the three TW
responses are guaranteed to hold.  Moreover, with the exception of
the ALM- ablation group (an exception Wicks himself has noted about
the ODE circuit model), our results match the relative-percentage
trends of Fig.~\ref{fig:bar_graph}.

Future work includes expanding the parameter ranges 
for TW responses, possibly by parallelizing the verification algorithm.
We also plan to examine the additional ablation groups present in 
Fig.~\ref{fig:bar_graph}.

\paragraph{Acknowledgments.}
We would like to thank Junxing Yang, Heraldo Memelli, Farhan Ali, 
and Elizabeth Cherry for their numerous contributions to this 
project.  Our research is supported in part by the following 
grants: NSF IIS 1447549, NSF CAR 1054247, 
AFOSR FA9550-14-1-0261, AFOSR YIP FA9550-12-1-0336, 
CCF-0926190, and NASA NNX12AN15H.

\bibliographystyle{abbrv}
\bibliography{cav15,mitra_cav2015}
\end{document}